REVIEW

# The Invention of Proteomic Code and mRNA Assisted Protein Folding.

by

*Jan C. Biro*

Homulus Foundation, 612 S Flower St. #1220, 90017 CA, USA.
jan.biro@att.net

**Keywords**:

Gene, code, codon, translation, wobble-base,

**Abbreviations** (excluding standard abbreviations:

## Abstract

**Background**

The theoretical requirements for a genetic code were well defined and modeled by George Gamow and Francis Crick in the 50-es. Their models failed. However the valid Genetic Code, provided by Nirenberg and Matthaei in 1961, ignores many theoretical requirements for a perfect Code. Something is simply missing from the canonical Code.

**Results**

The 3x redundancy of the Genetic code is usually explained as a necessity to increase the resistance of the mutation resistance of the genetic information.   However it has many additional roles. 1.) It has a periodical structure which corresponds to the physico-chemical and structural properties of amino acids. 2.) It provides physico-chemical definition of codon boundaries. 3.) It defines a code for amino acid co-locations (interactions) in the coded proteins. 4.) It regulates, through wobble bases the free folding energy (and structure) of mRNAs. I shortly review the history of the Genetic Code as well as my own published observations to provide a novel, original explanation of its redundancy.

**Conclusions**

The redundant Genetic Code contains biological information which is additional to the 64/20 definition of amino acids. This additional information is used to define the 3D structure of coding nucleic acids as well as the coded proteins and it is called the *Proteomic Code* and *mRNA Assisted Protein Folding*.

## Introduction

Mapping between messages in nucleic acid and protein alphabet is a fascinating story, a story that still unfolding. It is about to understand the rules of information transfer between DNA and proteins. First of all it is not only a biochemical puzzle and much of the early methods for devising codes came from combinatorics, information theory. Four and 20 (number of bases and amino acids) seems to be magical numbers with amazingly many possible mathematical connections between them [1].

The existence of a *Genetic Code* became obvious immediately after the discovery of DNA structure [2, 3]. The first suggestion for a code came from *George Gamow*, not even a biologist, but a physicist who became most famous as the chief proponent of the Big Bang theory in cosmology. He proposed a *Diamond Code* [4-6] where DNA acted directly as a template for assembling amino acids into proteins. Various combinations of bases along one of the grooves in the double helix could form distinctively shaped cavities into which the side chains of amino acids might fit. Each cavity would attract a specific amino acid; when all the amino acids were lined up in the correct order along the groove, an enzyme would come along to polymerize them. Gamow's code turned out to be an overlapping triplet code which provided exactly the desired 20 combinations. There are many beautiful aspects of the overlapping codon: a) it maximizes the density of information storage and b) even though three bases are needed to specify any single amino acids, the overall ratio of bases to amino acids approaches 1:1; c) it supposes that the distance between base pairs in DNA and the distance between amino acids in the proteins is raptly similar, which is exactly the case; 4) and avoids the possibility and consequences of frame shift. Unfortunately this code has serious constrains: only $4^4$ = 256 overlapping codon combinations are possible, while with 20 kinds of amino acids, there are $20^2$ = 400 possible dipeptides. The 144 "impossible" dipeptides were found in real proteins by *Sidney Brenner* [7] and it ruled out Gamow's codon "diamonds".

Another brilliant code was created by *Francis Crick* [8] the *Comma-Free Code*. Crick was speculating about an adaptor hypothesis (that he never published) the idea that amino acids do not interact directly with mRNA, but there is a mediator which recognizes the codons. Codons line up continuously along the DNA, like pearls on a neckless, and are recognized by a specific mediator. (This mediator is known today as tRNA). However, unlike a neckless, there is no space ("comma") between the codons which would indicate the codon boundaries. How is it still possible to distinguish between meaningful (those necessary 20 which lined up prettily on the DNA) and meaningless (the remaining 44 overlapping) codons? The answer was original: there is no mediator for meaningless codons. This solution was so simple and elegant, that it got the name the "prettiest wrong idea in all of 20[th]-century science".

There were many theoreticians involved in the invention of the Genetic Code. A common future of all theories was that they proposed some kind of connection between 20 amino acids and 20 codons. Nobody cared too much about biology. Crick wrote: "The importance of Gamow's work was that it was really an abstract theory of coding, and was not cluttered up by a lot of unnecessary chemical details..." But the same could be written about Crick. A historical curiosity is that not even Edward Teller, physicist and the father of the H-bomb, could resist giving away a theory about the Code.

Finally, *Marshall W Nirenberg* and *J Heinrich Matthaei*, two laboratory bench scientists published the real *Genetic Code* [9, 10] and provided a huge surprise: *it was redundant*. It was not compact, it was seemingly chaotic, there were no signs of any protection against frame shifts, and there were no signs of any connection between the codons and any characteristics of the coded amino acids. The real *Genetic Code* was boring, a real anticlimax, completely missing any intellectual elegance of a "big" scientific discovery. There is no esthetical beauty in the 64/20 code.

Why do we have 64 triplets for coding 20 amino acids - more than three times the number needed? Explaining away this excess became a major preoccupation of coding theorists. One intelligent explanation is that the redundancy confers a kind of error tolerance, in that many mutations convert between synonymous codons. When a mutation does alter an amino acid, the substitute is likely to have properties to those of the original. Alternatively the mutation is likely to be a stop codon which completely aborts the wrong translation. Another possible explanation is that the *Genetic Code* is developing, say started with 4, one letter codon coding 4 amino acids; continued to, say 16 two letter codons coding 16 amino acids. Recently we happen to have 64, three-letters-codons and they are coding 20 amino acids. However there is a potential to end up in the future with a, say 64, three-letters-codons coding 64 amino acids system.

Crick had, of course, his own explanation: forget any logical connection between codons and coded amino acids, it is just a "frozen accident" or by other words "that is what we got, lets we like it…"

## Second look at the Nirenberg Code

### *The 3D structure of mRNA*

Evolution often occurs in stages, one step is followed by a plateau before the next step is possible to take. It is true even for the development of scientific thinking and understanding. The 50-es and 60-es were very fertile for biology and the foundation of the recent molecular biology was laid. It took 30+ years to recognize, that the discovery of DNA was not the discovery of *The secret of Life*, the Life has much more secrets. Some early ideas were and some still are wrong, very wrong. The most embarrassing mistake of the modern genomics was probably the concept of sense and non-sense DNA strands. This was a concept deeply rooted in the mind of even the most brilliant scientists. The possibility of whole genome sequencing finally opened the gates (in late 90-es) to get rid of this nonsense idea and begin to see that both DNA strands are equally important for protein expression. It is no longer feasible, that one strand is expressed, while the other (complementary) strand serves only as a reproductive template (backup). Complementarity of bases is fundamental for helix formation of dsDNA, but more than that it makes possible formation of much more complex and sophisticated 3D structures than the known monotone helix. The unique, signature-structure of tRNA is well recognized and accepted, meanwhile the possibility and significance of mRNA 3D structure formation is widely denied [11]. The mRNA is passing the translation machinery on the surface of ribosomes, codon after codon, like a tape passes the magnets of a tape recorder. More linear mRNA is expected to be more suited for translation because a structured mRNA only slows down the machinery.

The number of mRNA images in Nucleic Acid Structure Database (NDB) is very limited [12]. Fortunately the known rules of codon base pairing makes it possible to predict the thermodynamically most likely structures of any nucleic acid [13]. (Such tools and predictions are not yet available for proteins). It is widely accepted today, that mRNAs have secondary structure even if there are numerous methodological considerations how to measure the folding energy content of nucleic acids [14]. Even if it is no longer possible to deny the existence of numerous significant secondary and tertiary structure elements in mRNA [15-16] their significance is not known and it is still remaining a disturbing element, especially for the biotech industry. The protein expression industry forced the development of tools and methods to speed up (so called "optimize") the expression velocity of proteins without altering its primary structure (the amino acid sequence). These methods are manipulating the wobble bases and replacing synonymous codons with each other to avoid structural elements in the entire sequence [17] or at least around critical codons, like the start codon [18].

### *Physico-chemical definition of codon boundaries*

Frame-shift is a major concern regarding the translation. Nirenberg's Genetic Code seems not to give any protection against the possible occurrence of frame-shifts even if it gives some promises to reduce the catastrophic consequences of wrong codon readings. However a second look at the base composition of codons (64 as it is), or the usage-weighed variants from different Species Specific Codon Usage Tables, reveals that it is not completely random. The first and third codon positions contain significantly more G and C bases, than the second (middle) codon positions. This bias has remarkable consequences. There are 3 H bonds between C and G (dG=-1524 kcal/1k bases) while only two between A and T bases (dG=-365 kcal/1k bases). Therefore the GC content is the major determinant of folding energy, and by that way the mRNA's 3D structure. The higher GC content indicates that the $1^{st}$ and $3^{rd}$ codon residues have significantly larger effect on the mRNA structure than the $2^{nd}$ one. Indeed, wobble bases were found to be the most important codon residues to determine mRNA structure [14].

This codon related periodic variation of GC content means that there is a periodic pattern of folding energy (dG) along the mRNA, which distinguishes the central codon base from the 1st and 3rd and forms a physico-chemical barrier or boundary between the codons. This is a statistical rule which doesn't apply for every single codon, but still shows a general tendency that there is some potential protection against frame shifts in the Nirenberg's *Genetic Code* [14, 19-20]. Manipulation of wobble bases to eliminate mRNA secondary structures (and speed up the translation) will destroy this energy pattern and by that way increase the translation errors. Native synonymous codon usage (and codon bias) seems to prefer selection for translational accuracy versus velocity [21-24].

### The Common Periodic Table of Codons and Amino Acids

There is even another completely separate line of evidence suggesting that codon positions are different, and the central codon position has a very special role. There always has been an effort to connect codons to their coded amino acids. The wobble base lost its importance because of its interchangeability. Most scientific efforts focused on to find stereo-chemical compatibility (spatial fitting) between the atomic geometry defined by 2 or 3 nucleic acid bases and the corresponding geometry defined by the residue of the coded amino acid [25, 26]. Crick furiously attached these efforts stating that any connection between codons and amino acids is only accidental and there is no underlying chemical rationale [27]. On the other hand *Carl Woese* argued that the *Genetic Code* developed in a way that was very closely connected to the development of the amino acid repertoire, and that this close biochemical connection is a fundamental of specific protein–nucleic acid interactions [28].

A common regularity in an arrangement of codons and amino acids provides a strong support for the evolutionary connection between mRNA and coded proteins.

A *Periodic Table of Codons* has been designed in which the codons are in regular locations. The *Table* has four fields (16 places in each), one with each of the four nucleotides (A, U, G, C) in the central codon position. Thus, AAA (lysine), UUU (phenylalanine), GGG (glycine) and CCC (proline) are positioned in the corners of the fields as the main codons (and amino acids). They are connected to each other by six axes. The resulting nucleic acid periodic table shows perfect axial symmetry for codons. The corresponding amino acid table also displaces periodicity regarding the biochemical properties (charge and hydropathy) of the 20 amino acids, and the positions of the stop signals. **Table I** emphasizes the importance of the central nucleotide in the codons, and predicts that purines control the charge while pyrimidines determine the polarity of the amino acids [29].

**Table I**

In addition to this correlation between the codon sequence and the physico-chemical properties of the amino acids, there is a correlation between the central residue and the chemical structure of the amino acids. A central *uridine* correlates with the functional group –C(C)$_2$–; a central *cytosine* correlates with a single carbon atom, in the C$_1$ position; a central *adenine* coincides with the functional groups –CC=N and –CC=O; and finally a central *guanine* coincides with the functional groups –CS, –C=O, and C=N, and with the absence of a side chain (glycine). (**Table II**)

I interpret these results as a clear-cut answer for the Woese vs. Crick dilemma: there *is* a connection between the codon structure and the properties of the coded amino acids. The second (central) codon base is the most important determinant of the amino acid property. It explains why the reading orientation of translation has so little effect on the hydropathy profile of the translated peptides. Note that 24 of 32 codons (U or C in the central position) code apolar (hydrophobic) amino acids, while only 1 of 32 codons (A or G in the central position) codes non-apolar (non-hydrophobic, charged or hydrophilic) amino acids. It explains why complementary amino acid sequences have opposite hydropathy, even if the binary hydropathy profile is the same.

**Table II**

**Effects of a Single Codon Residue
on the Structure of the Amino Acids**

***Visualization of specific nucleic acid – protein interactions***

The strong connection between codon structure and physicochemical properties of coded amino acids (the existence of *The Common Periodic Table of Codons and Amino Acids*) suggests that specific interaction between codons and coded amino acids might occur. There is no doubt, that specific interaction between nucleic acids and proteins is an absolute necessity for many vital functions, for example the regulation of gene expression. While should the codon / coded amino acid interaction be the only forbidden possibility to accomplish this function?

The interaction between restriction enzymes (RE) and their recognition sequences (RS) are known to be very specific and fortunately numerous such interactions are visualized and available from PDB. A review of the seven available crystallographic studies [30] showed that the amino acids coded by codons that are subsets of recognition sequences were often closely located to the RS itself and they were in many cases directly adjacent to the codon-like triplets in the RS. Fifty-five examples of this codon-amino acid co-localization were found and analyzed, which represents 41.5% of total 132 amino acids which are localized within 8 Å distance to the C1' atoms in the DNA. The average distance between the closest atoms in the codons and amino acids is 5.5 +/- 0.2 Å (mean +/- S.E.M, n = 55), while the distance between the nitrogen and oxygen atoms of the co-localized molecules is significantly shorter, (3.4 +/- 0.2 Å, p < 0.001, n = 15), when positively charged amino acids are involved. This is indicating that a direct interaction between nucleic- and amino acids might occur. We interpret these results in favor of Woese and suggest that the *Genetic Code* is "rational" and there is a stereo-specific relationship between the codons and the coded amino acids **(Figure 1).**

**Figure 1**
**Co-location of Codon-like Triplets and Amino Acids in RE-RS**

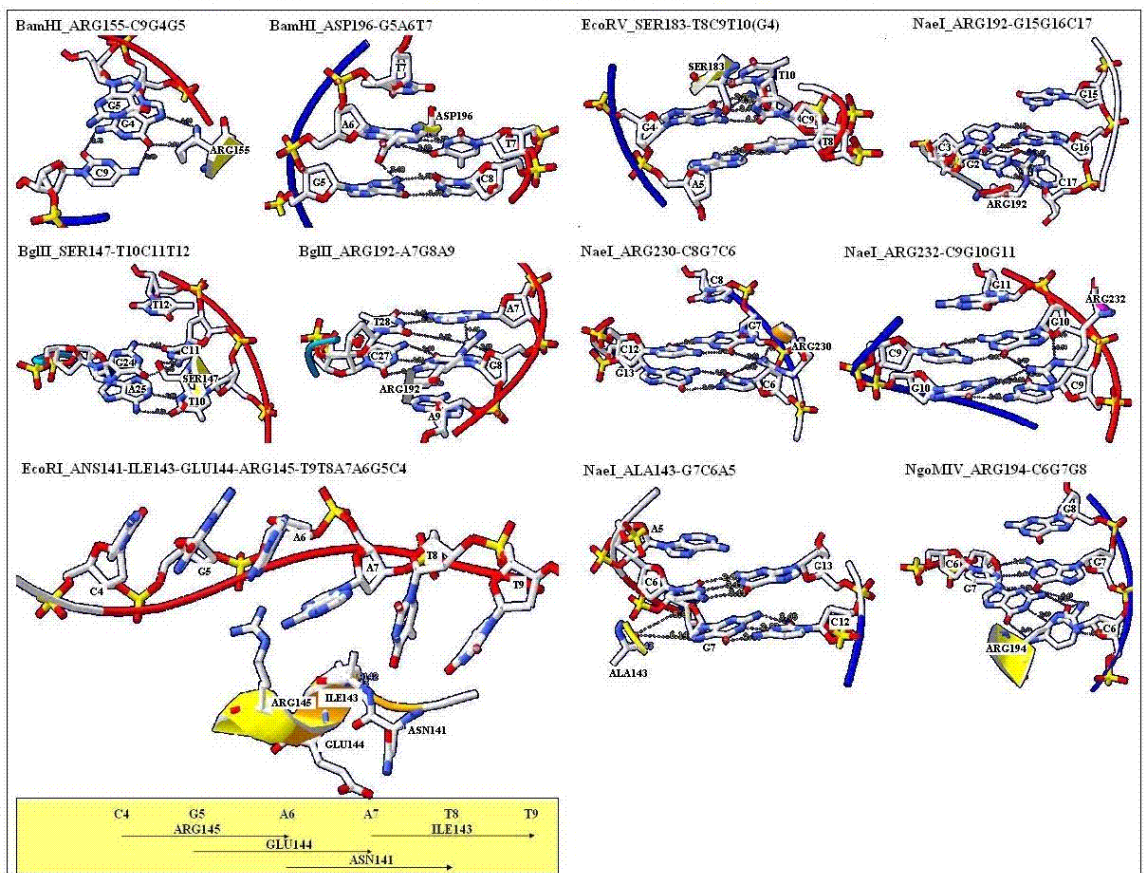

**Figure 1: Co-location of Codon-like Triplets and Amino Acids in RE-RS Complexes.**

Examples for co-locations of amino acids in restrictions endonucleases (RE) and codon-like triplets in restrictions enzyme recognition sites (RS). The name of enzyme, the name and position of nucleic acid bases and amino acids are indicated. Four amino acids located at overlapping-codon like base sequence in EcoRI is indicated in the yellow box.

### *Partial Complementary Coding of Co-locating Amino Acids.*

The observations above convinced me more and more that the connection between codons and coded amino acids (as it is described by Nirenberg) is not accidental and I started looking for an additional *"code in the code"* which might explain the redundancy of codons and give some rational explanation what the excess information in nucleic acids might be used for.

There was an idea published in early 80-s [31-35] suggesting that specifically interacting proteins are coded by complementary nucleic acids. The idea was, of course, rejected because it proposed that even the "non-sense" DNA strand might be expressed and makes some sense, which was an absurdity at that time. An early effort to confirm this theory, using (rather undeveloped) bioinformatical methods, failed [37]. There was a short come-back of this idea (called today as the *Proteomic Code*) and several research groups confirmed that proteins derived from complementary nucleic acid strands have specific, high affinity attraction to each other [38-42]. However it turned out that there is some problem with the consistency of the results: the method sometimes worked sometimes not.

We constructed a bioinformatics tool to collect data of co-locating amino acids from known protein structures, listed in the PDB, for statistical analyses [43]. (The immediate neighborhood on the same peptide chain was not counted as co-location). These analyses provided some rather novel observations.

1) Co-locating amino acids are physico-chemically compatible with each other, i.e. large and small, positive and negative, hydrophobe and hydrophobe amino acids are preferentially co-located with each other. The novelty of this observation is that physico-chemical rules apply already on residue level and do not necessarily need large, complex interfaces of interacting proteins [44].

2) Co-locating amino acids are preferentially coded by partially complementary codons, where the $1^{st}$ and $3^{rd}$ bases are complementary but the $2^{nd}$ may but not necessarily are complementary to each other (**Figure 2**).

**Figure 2**

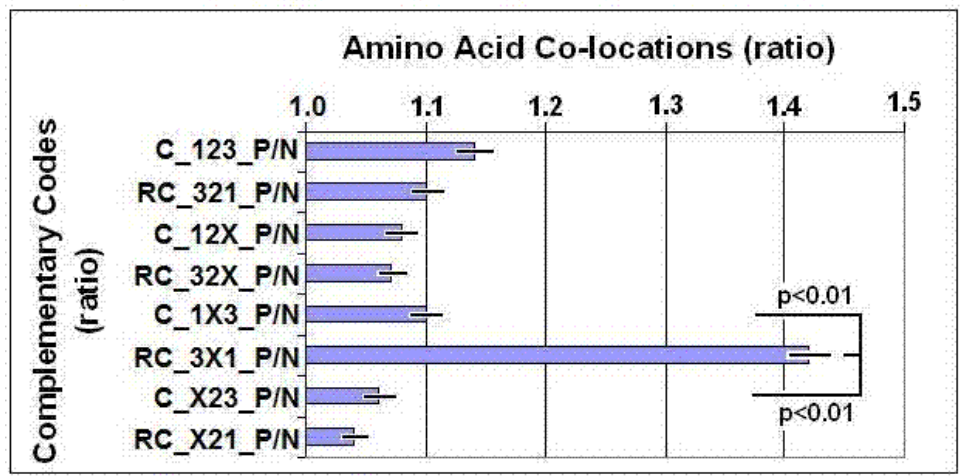

**Figure 2: Complementary codes vs. amino acid co-locations** (modified from [46]).

The propensities for the 20x20=400 possible amino acid pairs were monitored in 81 different protein structures with the SeqX tool. The tool detected co-locations when two amino acids were within 6 Å distance of each other (neighbors on the same strand were excluded). The total number of co-locations was 34,630. Eight different complementary codes were constructed for the codons (2 optimal and 6 suboptimal). In the two optimal codes, all three codon residues (123) were complementary (C) or reverse complementary (RC) to each other. In the suboptimal codes, only two of three codon residues were C or RC to each other (12, 13, 23), while the third was not necessarily complementary (X). (For example, Complementary Code RC_1X3 means that the first and third codon letters are always complementary, but not the second and the possible codons are read in reverse orientation.

These observations lead us to conclude that there are significant additional functional and structural connections between codons and coded amino acids to that what was described by Nirenberg and is known as the *Genetic Code*. We formulated the recent concept of *Proteomic Code* to describe this additional connection (for review see [45]).

### *The Role and Predictability of Wobble bases*

The 64/20 Genetic Code is redundant, mainly because the $3^{rd}$ codon bases, in most codons, are interchangeable without any consequence on the sequence of the coded protein. The only information expected from the DNA to the protein syntheses is the coding of amino acids, because it is believed, that the only information necessary to correct protein folding is only the correct amino acid sequence itself. These believe is based on *Anfinsen's thermodynamic principle* [47]. The Principle is more and more question-marked, both by bioinformation specialist and practicing protein scientist [48]. Some proteins are simply not folding toward one or a few thermodynamically well defined 3D destination and additional molecular information, provided by *chaperons*, is necessary to guide the protein folding.

By other words there is some 3x excess of information before translation, and there seems to be a shortage of information after translation. It is logical to assume, that folding information is stored in the redundant codons, more concretely in the wobble bases. The literature is actually rather rich with observations connecting the wobble bases to some structural feature of the coded proteins [49-54]. Even a wobble base centric sequence – structure database was constructed [55].

The preferential coding of co-locating amino acids by partially complementary nucleic acids, (for example by 5'>ANG>3'/3'<TNC<5' pattern) immediately suggests a role for the wobble bases. They are integrated parts of codons, defining amino acid co-locations. They are not randomly chosen, but logically selected: the wobble base of Xc codon (defining Xa amino acid) is defined by the first residue of codon Yc (which is coding Ya amino acid) if that two amino acids (Xa and Ya) are co-locating and *vice versa* the $3^{rd}$ residue of Yc is defined by the $1^{st}$ base in codon Xc (A defines T & U, G defines C).

Protein structures contain many amino acid co-locations (immediate neighbors on the same chain are excluded). Suppose that preferential partial complementarity coding of amino acids is not a rarity, but it is a rule. In that case the signs of non-randomness of wobble base selection should be seen not only in a small subset of proteins but even in very large data sets, like the species specific codon usage frequency tables.

Statistical analyses of A, T, G, C frequencies at $1^{st}$, $2^{nd}$, and $3^{rd}$ codon positions in 113 species specific Codon Usage Frequency Tables and 87 protein structures showed strongly significant internal correlation between the frequency of nucleic acid bases at different codon positions. This strong relationship made it possible to predict the frequency of all possible wobble bases in all the 64 codons in all the 113 species (P<1.3E-64, N=113) and all the 87 proteins (p<1.1E-28, n=87) [56].

These strong correlations wouldn't be possible with random selection of wobble bases. Therefore we concluded, that synonymous codons are not interchangeable with each other without disturbing the internal order of bases in integrated codon systems like a native mRNA or a species specific Codon Usage Frequency Order.

### Integrated Codon Systems

There are more than observations provided by theoretical and computational biology which are indicating, that native, natural proteins, as well as their coding sequences, are much more than the sequential collection of their building blocks. They are an integrated, interconnected system.

1) Wobble base mutations are expected to be "silent" without any consequences for the biological functions or phenotypes. They are often not. "Silent" polymorphism or mutation affects a) substrate specificity [57], b) drug pharmacokinetics and multidrug resistance in human cancer cells [58], c) mRNA stability and synthesis of the receptor [59], d) splicing [60, 61], e) different functions [62-64].

2) It has recently become clear that the classical notion of the random nature of mutations does not hold for the distribution of mutations among genes: most collections of mutants contain more isolates with two or more mutations than predicted by the mutant frequency on the assumption of a random distribution of mutations. Excesses of multiples are seen in a wide range of organisms, including riboviruses, DNA viruses, prokaryotes, yeasts, and higher eukaryotic cell lines and tissues. In addition, such excesses are produced by DNA polymerases *in vitro*. These "multiples" appear to be generated by transient, localized hypermutations rather than by heritable mutator mutations. The components of multiples are sometimes scattered at random and sometimes display an excess of smaller distances between mutations [65, 66].

3) A compensatory mutation occurs when the fitness loss caused by one mutation is remedied with a second mutation at a different site in the genome.

Often it occurs in the same gene, alters the protein sequence [67, 68] but saves the protein's secondary and tertiary structure. Compensatory mutations (in the same genes) seem to work through restoring the protein's structure (allosterism) which saves the protein's function [69].

Uneven concentration of mutations of smaller distances and their compensatory character are further indications of the integration and interconnectivity of codons in the same gene and consequently, conservation of structurally critical amino acid connections (but the amino acids) in the coded proteins.

### The RNA-assisted Protein Folding

The preferential partial complementarity coding of co-locating amino acids (*The Proteomic Code*) suggests the possibility that mRNA and coded proteins may share some common structural features. Side by side comparison of 2D projections of mRNA and coded proteins seems to confirm this possibility (**Figure 3, 4**).

### Figure 3

## Comparison of Protein and Corresponding mRNA Structures

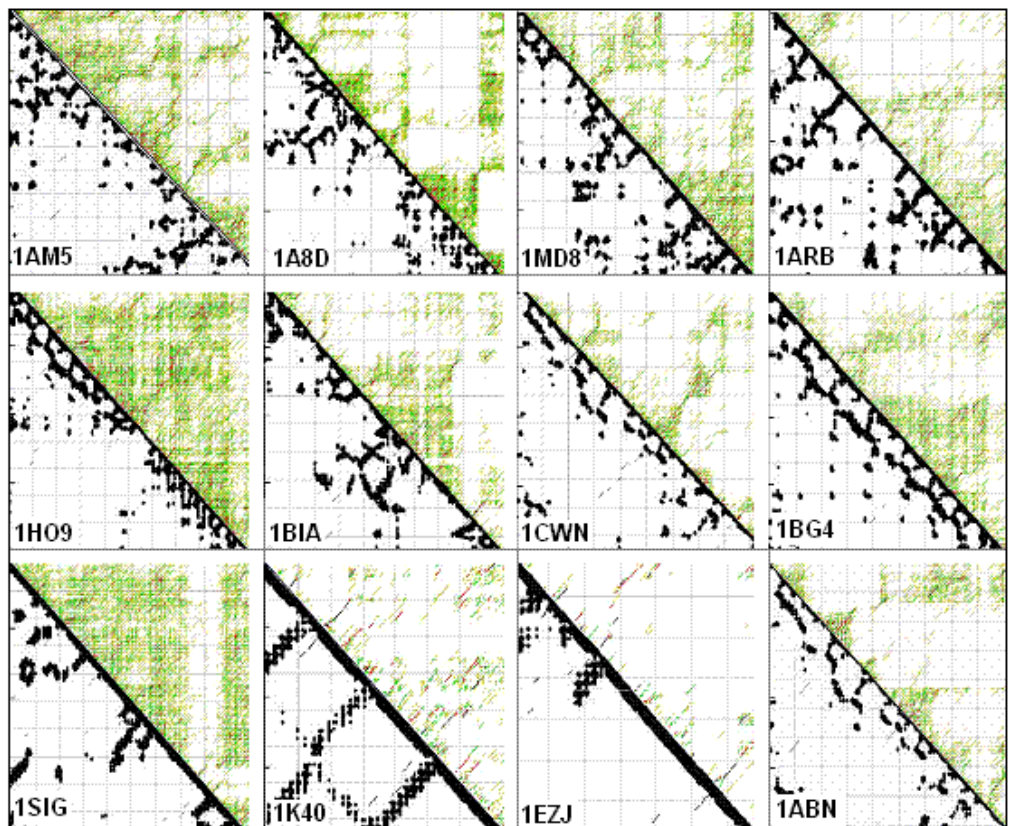

**Figure 3: Comparison of 12 randomly selected protein and corresponding mRNA structures** (modified from [46]).

Residue contact maps (RCM) were obtained from the PBD files of the protein structures using the SeqX tool (left triangles). Energy dot plots (EDP) for the coding sequences were obtained using the mfold tool (right triangles). The two maps were aligned along a common left diagonal axis to facilitate visual comparison between the different possible representations. The black dots in the RCMs indicate amino acids that are within 6 Å of each other in the protein structure. The colored (grass-like) areas in the EDPs indicate the energetically mostly likely RNA interactions (color code in increasing order: yellow, green red, black). The full names and the lengths of the proteins (number of amino acid residues): 1AM5: PEPSIN (324), 1A8D: TETANUS NEUROTOXIN (451), 1MD8: SERIN PROTEASE (329), 1ARB: ACHROMOBACTER PROTEASE I (268), 1HO9: A ALPHA-2A ADRENERGIC RECEPTOR (32), 1BIA: BIRA BIFUNCTIONAL PROTEIN (376), 1CWN: ALDEHYDE REDUCTASE (324), 1BG4: ENDO-1,4-BETA-XYLANASE (302), 1SIG: RNA POLYMERASE PRIMARY SIGMA FACTOR (339) bases, 1K40: ADHESION KINASE (126), 1EZJ: NUCLEOCAPSID PHOSPHOPROTEIN (140), 1ABN: ALDOSE REDUCTASE (315). The coordinates indicate the number of amino acid and the corresponding nucleic acid residues.

## Figure 4
## Comparison of the Protein and mRNA Secondary Structures

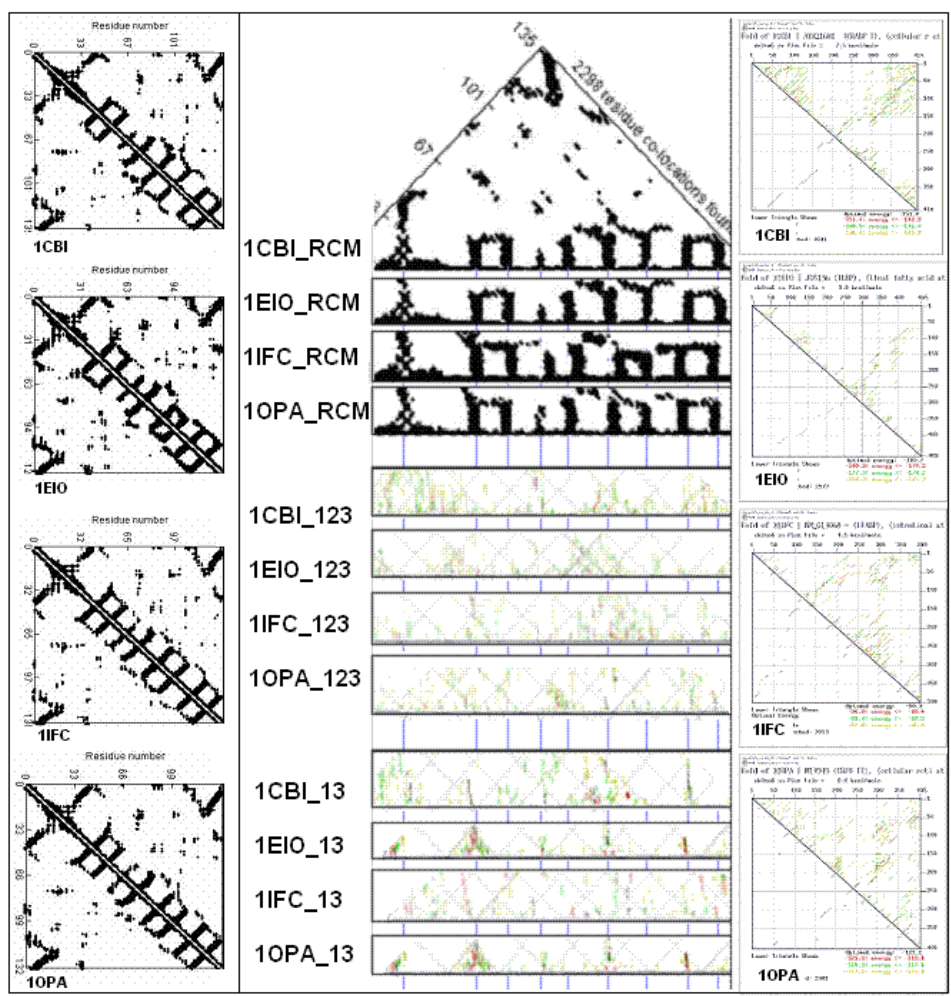

**Figure 4: Comparison of the protein and mRNA structures** (modified from [46]).

Residue contact maps (RCM) were obtained from the PBD files of four protein structures (1CBI, 1EIO, 1IFC, 1OPA) using the SeqX tool (left column). Energy dot plots (EDP) for the coding sequences were obtained using the mfold tool (right column). The left diagonal portions of these two maps are compared in the central part of the figure. Blue horizontal lines in the background correspond to the main amino acid co-location sites in the RCM. Intact RNA (123) as well as subsequences containing only the 1st and 3rd codon letters (13) are compared. The black dots in the RCMs indicate amino acids that are within 6 Å of each other in the protein structure. The colored (grass-like) areas in the EDPs indicate the energetically most likely RNA interactions (color code in increasing order: yellow, green red, black). The full names and the lengths of the proteins (number of amino acid residues): 1CBI: CELLULAR RETINOIC ACID BINDING PROTEIN I (136), 1EIO: ILEAL LIPID BINDING PROTEIN (127), 1IFC: INTESTINAL FATTY ACID BINDING PROTEIN (132), 1OPA: CELLULAR RETINOL BINDING PROTEIN II (135).

Biological rules are, of course, always statistical rules, probabilities and tendencies. Nucleic acids as well as proteins have many possible configurations where one or a few are expected to dominate and define the main and characteristic configuration. Coding- and coded sequences might have their own range of more or less different folding potentials. However when a protein is generated on the surface of ribosome the coding- and coded sequences are very close to each other. This temporary intimate closeness is a possibility for coding sequences for transfering folding information to coded proteins, information that is additional to that these proteins already have in their amino acid sequences. Some ideas how it is possible are sketched in **Figure 5 and 6**).

**Figure 5**
**RNA Assisted Protein Loop Formation**

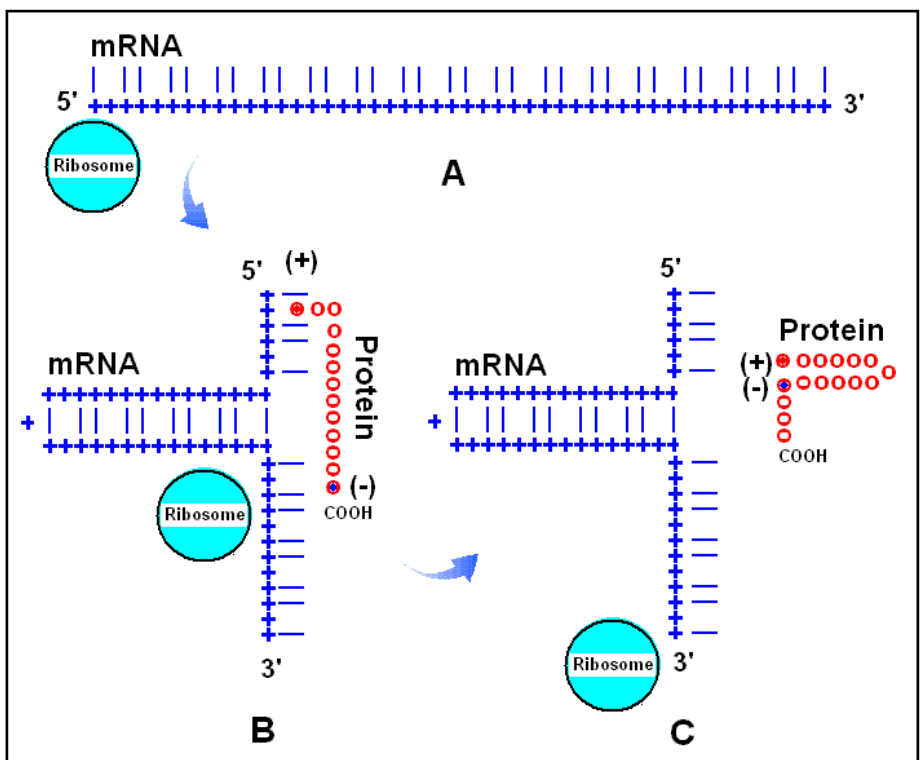

**Figure 5: RNA assisted protein loop formation**
Translation begins with the attachment of the 5' end of a mRNA to the ribosome (A). Ribonucleotides are indicated by blue + and the 1st and 3rd bases in the codons by blue lines, while the 2nd base positions are left empty. A positively charged amino acid [(+) and red dots], for example arginine, remains attached to its codon. The mRNA forms a loop because the 1st and 3rd bases are locally complementary to each other in reverse orientation (B). The growing protein is indicated by red circles (o). When translation proceeds to an amino acid with especially high affinity to the mRNA-attached arginine, for example a negatively charged Glu or Asp [(-) and blue dot], the charge attraction removes the Arg from its mRNA binding site and the entire protein is released from the mRNA and completes a protein loop (C). The protein continues to grow toward the direction of its carboxy terminal (COOH). (Figure is reproduced from [70]).

## Figure 6

### RNA-Assisted (translational) Protein Folding

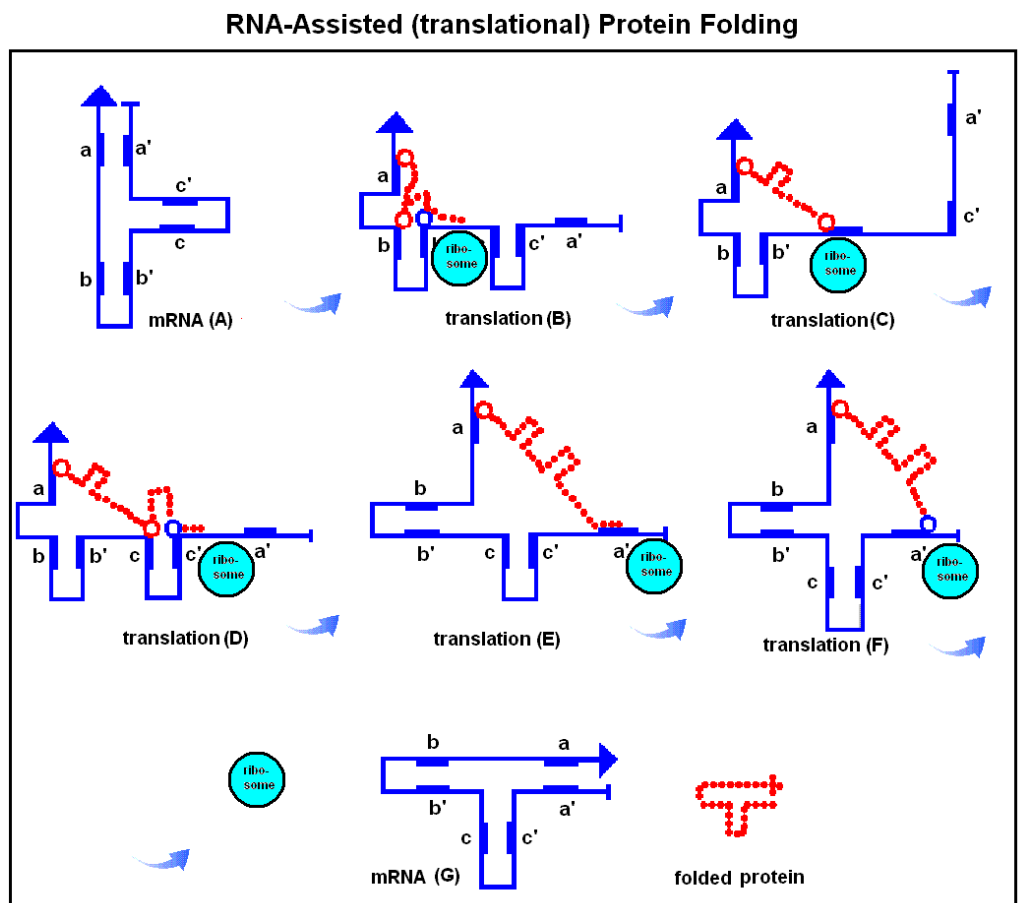

**Figure 6: RNA-assisted (translational) protein folding.**

There are three reverse and complementary regions in a mRNA (blue line, A): a-a', b-b', c-c', which fold the mRNA into a T-like shape. During the translation process the mRNA unfolds on the surface of the ribosome, but subsequently refolds, accompanied by its translated and lengthening peptide (red dotted line, B-F). The result of translation is a temporary ribonucleotide complex, which dissociates into two T-shape-like structures: the original mRNA and the properly folded protein product (G). The red circles indicate the specific, temporary attachment points between the RNA and protein (for example a basic amino acid) while the blue circles indicate amino acids with exceptionally high affinity for the attachment points (for example acidic amino acids); these capture the amino acids at the attachment point and dissociate the ribonucleoprotein complex. Transfer-RNAs are of course important participants in translation, but they are not included in this scenario. (Figure is reproduced from [70]).

Transfer RNA (tRNA) is not to forget. There is a feature of this molecule which indicates that there is more to expect from these special kinds of nucleic acids than only transfer amino acids to the proper codons on the mRNA. They are too large. Codon recognition, amino acid recognition, amino acid binding and transfer could be achieved by much smaller and less complex molecules. The energy cost to synthesize and maintain these big, bulky molecular structures is not proportional to its known, simple function. This indicates for me that more function is to expect from tRNAs.

### Evolutionary Aspects of the Redundant Genetic and Proteomic Codes.

The *Common Periodic Table of Codons and Amino Acids* is convincing evidence for the non-randomness of codon / coded amino acid connection. This connection is further emphasized by the preferential complementary coding of co-locating amino acids. The predictability of wobble base frequency is another indicator of non-random wobble base selection and a functional network between codons. Order is traditionally seen as the result of development from the chaotic to organized, from simple to complex. This process is called evolution.

The theory of Ikehara [71, 72] about the origins of gene, genetic code, protein and life is especially interesting regarding the *Proteomic Code*. Ikehara suggests (and support with experimental evidence) that "geneprotein" system, comprised of 64 codons and 20 amino acids developed successfully during the evolution. The development started with a GNC-type primeval genetic code (G: guanine, C: Cytosine, N: any of the four nucleotides), coding only four amino acids (Gly: [G], Ala: [A], Asp: [D], Val: [V]) forming the so called [GADV]-proteins. This minimal set of only four amino acids and the [GADV]-proteins are able to represent the 6 major (and characteristic) protein moieties/indices (hydropathy, a helix, b-sheet and b-turn forms, acidic amino acid content and basic amino acid content) which are necessary for appropriate three-dimensional structure formation of globular, water-soluble proteins on the primitive earth. The [GADV]-proteins (even randomized) have catalytic properties and able to facilitate the syntheses of other [GADV]-proteins (also random).

The primeval genetic code continued to develop toward a more complex SNS-type primitive genetic code (S: G or C) containing 16 codons and encoding 10 amino acids (L, P, H, Q, R, V, A, D, E, G) before the recent 64 codon/20 amino acid-type recent genetic code became established.

Furthermore, Ikehara concluded from the analysis of microbial genes that newly-born genes are products of nonstop frames (NSF) on antisense strands of microbial GC-rich genes [GC-NSF(antisense)] and from SNS repeating sequences [(SNS)n] similar to the GC-NSF(antisense).

The similarity between GNC/SNS-type primitive codons (which are expressed even from the reverse-complement strands as GC-rich non-stop genes) and the *Proteomic Code* is obvious. Both concepts suggest and agree with each other regarding a) the connection between 2nd codon residue and the fundamental physicochemical properties of the coded amino acids, b) the importance of 1st and 3rd codon letters in determining the nucleic acid (as well as protein) structure, c) the importance of compositional difference between 1st, 3rd and central codon residues (to emphasize the codon boundaries), d) the importance of complementarity (even in the mRNA) in development of protein structure and function, e) the importance of GC at the 1st and 3rd codon positions (as the source of lower Gibbs energy (dG), than central codon positions have, where even AT are permitted). I think that the concept of GNC/SNS-type primitive codons and the Proteomic Code are convergent ideas, both reflecting the same fundamental aspects of the connection between nucleic acid and protein structure and function.

## Conclusions

The Nirenberg's *Code* seems to have another face that was in the shadow some 40+ years. We start to see, that the *Genetic Code* actually might contain all characteristics that was expected and promised by the early theoretical code models.

1) Codon boundaries are physico-chemically defined to a certain degree, which theoretically should give some protection against frame-shifts.

2) Codon residues are not randomly assigned, but there is a connection between codon architecture and the physicochemical properties of the coded proteins.

3) Amino acids preferentially interact with their codons (studied in restrictions endonucleases).

4) Co-locating amino acids are preferentially coded by partially complementary codons which create inter-connectivity between structurally important amino acids.

5) Wobble bases are not randomly assigned at all, their frequency is statistically well predictable from the frequency of bases at other codon positions.

6) Wobble base redundancy makes it possible the development of codon integration in coding sequences which might be used for compensatory mutations. This is the second line of defense against mutations (after the known tolerance provided by the coding redundancy).

7) The internally inter-connected and integrated system of codons makes it possible that coding sequences provide a mold for structure forming of coded proteins and function as *nucleic acid chaperons* providing the missing molecular information to correct protein folding.

It is concluded that the redundant *Genetic Code* contains biological information which is additional to the 64/20 definition of amino acids. This additional information is used to define the 3D structure of coding nucleic acids and coded proteins and is called the *Proteomic Code*.

After this second look at the Nirenberg's Code it is not at all "gray" and "boring" anymore and seems to be on a good way to provide even the expected intellectual beauty of a "big" discovery.